%% file: main_sigkdd.tex
\documentclass[sigconf]{acmart}

\input{macro}

\AtBeginDocument{%
  }

\setcopyright{acmlicensed}
\copyrightyear{2018}
\acmYear{2018}
\acmDOI{XXXXXXX.XXXXXXX}
\acmConference[Conference acronym 'XX]{Make sure to enter the correct
  conference title from your rights confirmation email}{June 03--05,
  2018}{Woodstock, NY}
\acmISBN{978-1-4503-XXXX-X/2018/06}

\begin{document}

%%
%% The "title" command has an optional parameter,
%% allowing the author to define a "short title" to be used in page headers.
\title{TaxoConf: Taxonomy-Guided Automatic \TaskName{}}

%%
%% The "author" command and its associated commands are used to define
%% the authors and their affiliations.
%% Of note is the shared affiliation of the first two authors, and the
%% "authornote" and "authornotemark" commands
%% used to denote shared contribution to the research.
% \author{Ben Trovato}
% \authornote{Both authors contributed equally to this research.}
% \email{trovato@corporation.com}
% \orcid{1234-5678-9012}
% \author{G.K.M. Tobin}
% \authornotemark[1]
% \email{webmaster@marysville-ohio.com}
% \affiliation{%
%   \institution{Institute for Clarity in Documentation}
%   \city{Dublin}
%   \state{Ohio}
%   \country{USA}
% }

% \author{Daomin Ji}
% \affiliation{%
%   \institution{RMIT University}
%   }
% \email{daomin.ji@student.rmit.edu.au}

% \author{Hui Luo}
% \affiliation{%
%   \institution{University of Wollongong}
%   }
% \email{huil@uow.edu.au}

% \author{Zhifeng Bao}
% \affiliation{%
%   \institution{The University of Queensland}
%   }
% \email{zhifeng.bao@uq.edu.au}

% \author{Zi Huang}
% \affiliation{%
%   \institution{The University of Queensland}
%   }
% \email{huang@itee.uq.edu.au}

% \author{A. K. Qin}
% \affiliation{%
%   \institution{Swinburne University of Technology}
%   }
% \email{kqin@swin.edu.au}

% \author{Daomin Ji$^{1}$, Zhifeng Bao$^{1}$, Junhao Gan$^{2}$}

% \affiliation{$^{1}$The University of Queensland; $^{2}$The University of Melbourne}

% \email{
% d.ji@uq.edu.au; zhifeng.bao@uq.edu.au; junhao.gan@unimelb.edu.au
% }

\author{Daomin Ji}
\affiliation{%
  \institution{The University of Queensland}
}
\email{d.ji@uq.edu.au}

\author{Zhifeng Bao}
\affiliation{%
  \institution{The University of Queensland}}
\email{zhifeng.bao@uq.edu.au}

\author{Junhao Gan}
\affiliation{%
  \institution{The University of Melbourne}}
\email{junhao.gan@unimelb.edu.au}

% \author{John Smith}
% \affiliation{%
%   \institution{The Th{\o}rv{\"a}ld Group}
%   \city{Hekla}
%   \country{Iceland}}
% \email{jsmith@affiliation.org}

% \author{Julius P. Kumquat}
% \affiliation{%
%   \institution{The Kumquat Consortium}
%   \city{New York}
%   \country{USA}}
% \email{jpkumquat@consortium.net}

%%
%% By default, the full list of authors will be used in the page
%% headers. Often, this list is too long, and will overlap
%% other information printed in the page headers. This command allows
%% the author to define a more concise list
%% of authors' names for this purpose.
\renewcommand{\shortauthors}{Ji et al.}

%%
%% The abstract is a short summary of the work to be presented in the
%% article.
\input{sections/sec_0_abstract}

\received{20 February 2007}
\received[revised]{12 March 2009}
\received[accepted]{5 June 2009}

\maketitle

\input{sections/sec_1_introduction}
\input{sections/sec_2_problem_formulation}
\input{sections/sec_3_solution_part_1}
\input{sections/sec_4_solution_part_2}

\input{sections/sec_5_solution_part_3}
\input{sections/sec_6_experiment}

\input{sections/sec_7_deployment}
\input{sections/sec_8_related_work}
\input{sections/sec_9_conclusion}

%\clearpage

\bibliographystyle{ACM-Reference-Format}
\bibliography{reference}

\input{sections/sec_appendix}
\end{document}

%% file: macro.tex
\PassOptionsToPackage{a4paper,margin=1in}{geometry}

\usepackage{xspace}
\usepackage{ifthen}
\usepackage{marginnote}
\usepackage{booktabs,tabularx,multirow}

\usepackage{geometry}  
\usepackage{microtype}
\usepackage{times}

\usepackage{graphicx}
\usepackage{caption}
\usepackage{enumitem}
\usepackage{xcolor}

\usepackage{mathtools,amsfonts,bm}
\usepackage{amsthm}

\usepackage{hyperref}
\usepackage{titlecaps}

\usepackage{graphicx}
\usepackage{subcaption}

\usepackage{algorithm}
\usepackage{algpseudocode}

\newtheorem{definition}{Definition}

\makeatletter
\g@addto@macro\normalsize{%
  \setlength{\abovedisplayskip}{0pt}%
  \setlength{\abovedisplayshortskip}{0pt}%
  \setlength{\belowdisplayskip}{0pt}%
  \setlength{\belowdisplayshortskip}{0pt}%
}

\makeatother

\makeatletter
\def\thm@space@setup{%
  \thm@preskip=2pt
  \thm@postskip=2pt
}
\makeatother
\newcommand{\Papers}{\mathcal{P}}
\newcommand{\Sessions}{\mathcal{S}}
\newcommand{\Tracks}{\mathcal{T}}

\newcommand{\Fcoh}{F_{\mathrm{coh}}}

\newcommand{\Front}{\mathcal{F}}

\newcommand{\taskname}{conference program organization\xspace} % mid-sentence
\newcommand{\Taskname}{Conference program organization\xspace} % sentence start
\newcommand{\TaskName}{Conference Program Organization\xspace} % titles/headings

\newcommand{\sys}{TaxoConf\xspace}
\newcommand{\TopicILP}{Topic-ILP\xspace}
\newcommand{\EmbILP}{Embedding-ILP\xspace}
\newcommand{\LLMDirect}{LLM-Direct\xspace}
\newcommand{\LLMReAct}{LLM-ReAct\xspace}

%% file: sections/sec_0_abstract.tex
\begin{abstract}
\Taskname{}, the task of assembling accepted papers into a technical program, is labor-intensive. Papers must be grouped into topically coherent sessions under hard operational constraints, and existing methods rarely achieve both at once. To address this problem, we present \sys{}, a system that organizes conference programs around a conference-specific topic taxonomy. \sys{} constructs a canonicalized multi-parent taxonomy over the accepted papers and represents each paper by its frontier of most specific topics. It derives a specificity-weighted optimal-transport distance between papers from this taxonomy, then solves a binary integer programming problem that assigns papers to sessions by minimizing within-session distance subject to all hard constraints. On benchmarks built from four 2025 conferences, \sys{} attains the highest session coherence (4.62 out of 5) and the closest agreement with human-curated sessions (NMI 0.761) while incurring no constraint violations. \sys{} has also been deployed to generate the technical program of SIGIR~2026, where the organizers accepted the initial output with only a few edits, preserving 91.8\% of oral assignments and the entire poster program while requiring no correction of hard-constraint violations, and an on-site survey of 47 attendees rated the overall program quality 4.26 out of 5. The system is publicly available at \url{https://taxoconf.com}.
\end{abstract}

%% file: sections/sec_1_introduction.tex
\section{Introduction}
\label{sec:introduction}

\Taskname{} is a recurring task across the academic community: each year, many conferences must arrange hundreds to thousands of accepted papers into a coherent technical program. Conference management platforms such as EasyChair, Microsoft CMT, and OpenReview streamline submission and peer review \cite{easychair,microsoftcmt,openreview}, and a large body of work automates reviewer assignment \cite{charlin2013tpms,stelmakh2021peerreview4all,leytonbrown2024matching}. However, the downstream step of assembling accepted papers into a session-based program remains largely manual. Chairs read titles and abstracts, group related papers into sessions, name each session, and place sessions into rooms and time slots subject to operational constraints. For example, the organizers of SIGIR 2026 had to arrange 657 accepted presentations into 51 oral sessions and 3 poster sessions while satisfying a set of hard constraints (see Sec.~\ref{sec:deployment} for details). This manual process consumes considerable time and effort, motivating automated support for \taskname{}.

The task is challenging because the resulting program should satisfy two properties that pull in different directions. \textbf{(P1) Topical coherence.} Each session should group papers that address related problems, methods, or applications, so that attendees can follow the session as a \textit{thematic unit}. \textbf{(P2) Operational constraints.} The program must satisfy all hard constraints imposed by organizers, presenters, and available resources, such as session capacity bounds, track compatibility, and presenter conflicts across parallel sessions. Semantic grouping alone does not yield a deployable program, and constraint satisfaction alone does not yield a coherent one, so a method must achieve both simultaneously rather than optimize one and repair the other.

Many methods have been proposed to reduce this burden, ranging from interactive assistance to automated \taskname{}. Human-in-the-loop systems elicit paper-affinity judgments from the broader community or the program committee and use them to support interactive session construction~\cite{kim2013cobi,andre2013community,chilton2014frenzy}. Although effective, these systems require substantial additional input from participants and organizers, so the overall human effort is not substantially reduced and scalability remains limited. Optimization-based methods instead formulate the task as an assignment problem and explicitly enforce operational constraints~\cite{sidiropoulos2015signal,bulhoes2022conference}. However, they typically measure topical relatedness by flat pairwise similarities derived from topic models~\cite{blei2003lda} or text embeddings~\cite{cohan2020specter}. Such similarities capture local semantic proximity but do not represent the hierarchical and cross-cutting relationships among research topics, which limits the topical coherence of the resulting sessions. More recently, LLMs have been explored for \taskname{}~\cite{jobson2024investigating}, motivated by their demonstrated ability to discover complex topic structures from scientific document collections~\cite{pham2024topicgpt,hsu2024chime,zeng2024chainoflayer}. Through direct prompting or multistep reasoning strategies such as chain-of-thought prompting~\cite{wei2022chain}, an LLM can in principle produce a topically coherent program. Yet such prompting-based approaches lack a mechanism to enforce conference-wide operational constraints. 

\begin{figure*}[th]
    \centering
    \includegraphics[width=\linewidth]{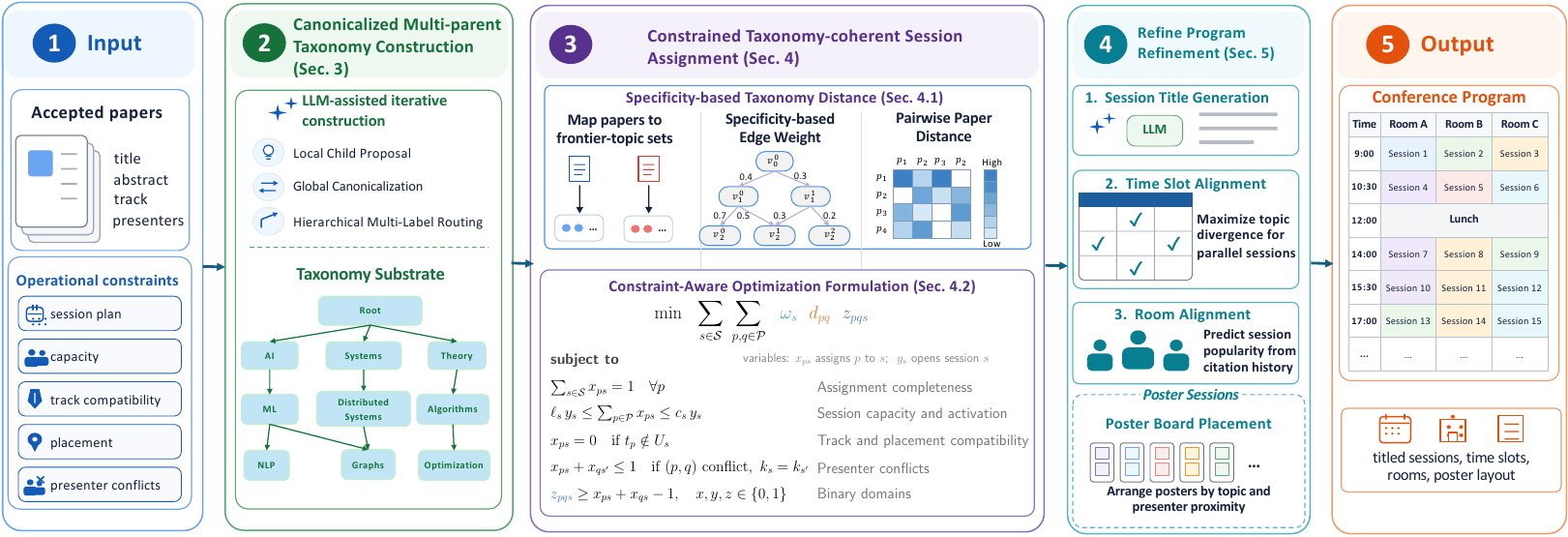}
    \caption{The workflow of \sys.}
    \label{fig:workflow}
\vspace{-2em}
\end{figure*}

To address these limitations, we present \sys{}, whose core design principle is to separate semantic modeling from combinatorial scheduling, as illustrated in Figure~\ref{fig:workflow}. The main technical effort lies in the semantic modeling, which answers two questions: how to build a faithful representation of the topic structure of the accepted papers, and how to derive from it a distance that accurately measures the semantic relation between any two papers. For the representation, we propose a novel LLM-based taxonomy construction method. Unlike existing LLM-based approaches, which often struggle to maintain structural consistency~\cite{pham2024topicgpt,hsu2024chime,zeng2024chainoflayer}, our method ensures three structural properties: \emph{global canonicality} (each topic appears as a single shared node rather than being duplicated across branches), \emph{polyhierarchy} (a topic may have multiple parents, so cross-cutting topics spanning several research areas are represented, making the taxonomy a DAG rather than a tree), and \emph{strict hierarchy} (every child topic is strictly more specific than each of its parents). For the distance, \sys{} weights the taxonomy edges by information content~\cite{resnik1995ic,lin1998similarity} and measures the distance between two topics by the shortest path length under these weights. Each paper is treated as a uniform distribution over its frontier topics, and the distance between two papers is the optimal-transport cost between these distributions, where moving mass from one topic to another costs their topic distance~\cite{peyre2019computational,rubner2000emd}. For the combinatorial scheduling, \sys{} casts session assignment as a mixed-integer linear program: the objective minimizes the weighted within-session distance between papers, all hard operational constraints are encoded as linear constraints, and the resulting program is solved using off-the-shelf solvers~\cite{gurobi,ortools,scip}. Finally, a post-assignment refinement stage generates session titles using an LLM grounded in the aggregated taxonomy evidence, schedules topically divergent sessions in parallel time slots, and matches sessions to rooms using citation-based popularity estimates.

In summary, we make the following contributions.
\begin{itemize}[leftmargin=*,noitemsep,topsep=0pt]
\item We present \sys{}, a novel end-to-end framework that constructs topically coherent conference programs while satisfying real-world operational constraints. (Sec.~\ref{sec:problem}).
\item We propose a globally canonicalized multi-parent taxonomy construction method that builds an interpretable topic DAG, captures hierarchical and cross-cutting relationships, and represents each paper by its most specific topic frontier (Sec.~\ref{sec:taxonomy-construction}).
\item We develop a taxonomy-guided paper distance based on information content and optimal transport and incorporate it into a constraint-aware integer programming formulation for the \taskname{} task (Sec.~\ref{sec:optimization-framework}).
\item We construct four new benchmarks from the official programs of major 2025 conferences and define a comprehensive evaluation protocol covering semantic coherence, agreement with human-curated sessions, and constraint satisfaction. Experiments show that \sys{} achieves the best coherence and human agreement while incurring zero constraint violations (Sec.~\ref{sec:experiments}).
\item We present the real-world deployment of \sys{} at SIGIR 2026: the final program preserved 91.8\% of \sys{}'s initial oral paper-to-session assignments, adopted the generated poster program without modification, and required no correction of hard-constraint violations. An on-site survey of 47 attendees rated the overall program quality 4.26 out of 5. \sys{} is further slated for adoption by upcoming conferences, including VLDB~2026, SIGIR~2027, ICDE~2027, and DASFAA~2027 (Sec.~\ref{sec:deployment}).
\end{itemize}

%% file: sections/sec_2_problem_formulation.tex
\section{Problem Formulation}
\label{sec:problem}

We are given a collection of accepted papers $\Papers$ to be organized into a conference program. Each paper $p\in\Papers$ has textual content and associated metadata. For the textual content, we primarily use the title and abstract, which are generally sufficient to identify topical relationships among papers. The metadata include the presenters and the track $t_p\in\Tracks$ to which the paper belongs, both of which are central to the operational constraints.

We assume that the organizers have specified a session plan capable of accommodating the accepted papers. Let $\Sessions$ denote the set of candidate sessions. Each session $s\in\Sessions$ is characterized by its room, time slot, capacity bounds, and track requirements. We write $s=(r_s,k_s,\ell_s,c_s,U_s)$, where $r_s$ is the room, $k_s$ is the time slot, $\ell_s$ and $c_s$ are the minimum and maximum numbers of papers that the session may contain, and $U_s\subseteq\Tracks$ is the set of permitted tracks. Thus, $U_s$ specifies whether a session may mix papers from multiple tracks, such as the research and applied data science tracks, or is restricted to a single track.

Program construction is governed by two complementary goals that jointly determine the quality and validity of the resulting program but may conflict in practice.
\begin{itemize}[noitemsep,leftmargin=*,topsep=0pt]
\item \textbf{Topical coherence.} Papers assigned to the same session should address closely related problems, methods, or applications, allowing the session to form a meaningful thematic unit.
\item \textbf{Operational constraints.} The program must satisfy all mandatory operational requirements imposed by the organizers, presenters, and available resources, such as presenter availability, session-capacity limits, and track compatibility.
\end{itemize}

\smallskip
We formally define the task as follows.

\begin{definition}[\TaskName{}]\label{def:problem}
Given a set of accepted papers $\Papers$ and a set of candidate sessions $\Sessions$, let $\sigma:\Papers\rightarrow\Sessions$ denote a paper-to-session assignment and let $\Fcoh(\sigma)$ denote its topical-coherence score, with larger values indicating greater coherence. Let $\Omega_{\mathrm{op}}$ denote the set of assignments that place every paper in exactly one session and satisfy all mandatory operational requirements. \Taskname{} seeks an assignment
$\sigma^{*}\in\arg\max_{\sigma\in\Omega_{\mathrm{op}}}\Fcoh(\sigma)$,
thereby maximizing topical coherence while maintaining operational feasibility.
\end{definition}

%% file: sections/sec_3_solution_part_1.tex
\section{Canonicalized Multi-Parent Taxonomy Construction}
\label{sec:taxonomy-construction}

\input{algs/taxonomy_construction}

The taxonomy is the central representation that drives paper grouping, so its structure directly affects the quality of the resulting program. In an academic taxonomy, a single topic may specialize more than one broader topic, so a node may have multiple parents and the taxonomy should form a directed acyclic graph (DAG) rather than a tree. Most existing LLM-based taxonomy construction methods instead expand the taxonomy iteratively, recursively dividing papers into groups, which yields a tree and cannot express this relation~\cite{wan2024tntllm, zeng2024chainoflayer, kargupta2025taxoadapt}.
Building an accurate and faithful DAG over accepted papers is challenging. Because branches are expanded independently, nodes discovered in different branches may denote the same topic, leaving equivalent nodes scattered across the taxonomy. What aggravates this is that branches may expand to different depths and semantic granularities, so topics at the same nominal level need not be comparable in specificity. Both issues degrade the downstream paper assignment, since a paper's position in the taxonomy then depends on which branch happened to reach it rather than on its actual topical content.

To address these challenges, we propose \emph{canonicalized multi-parent taxonomy construction}, which uses an LLM to iteratively divide the papers into finer topics and resolves every proposed topic against a global registry, so that equivalent topics discovered across branches collapse onto shared nodes and a node may acquire multiple parents. In each iteration, the routing of a paper to the proposed subtopics is treated as a multi-label classification, so a paper may follow several branches simultaneously. Algorithm~\ref{alg:taxonomy-construction} summarizes the full procedure, and we elaborate each step below.

\noindent\textbf{Initialization} (Line~\ref{ln:init}).
We start construction from a single root topic because the accepted papers initially form one undifferentiated corpus. The root provides a common ancestor for all subsequent topics and makes upward closure well-defined for every paper. The registry $\mathcal{V}$ is initialized with the root alone, every paper is assigned to $r$, and the expansion queue $\mathcal{Q}$ is seeded with $r$.

\noindent\textbf{Local Child Proposal} (Lines~\ref{ln:pop}, \ref{ln:propose}--13).
At each iteration, \sys{} dequeues a topic node $v$ from the queue $\mathcal{Q}$ and expands it into a set of child nodes $C(v) = \{v_1, \dots, v_k\}$, where each child $v_i$ carries a topic name and a description that together characterize a subtopic of $v$. Following existing methods~\cite{kargupta2025taxoadapt}, we prompt the LLM to propose candidate child topics as $C(v) = \mathrm{LLM}_{\text{propose}}\big(D(v),\, \langle \mathrm{name}(v), \mathrm{def}(v) \rangle\big)$, conditioned on two inputs: (i) the papers currently contained in $v$, denoted $D(v)$ and represented by their titles and abstracts, and (ii) the node itself, described by its canonical name and definition. Here $D(v)$ supplies the corpus evidence for which subtopics are present, and $\langle \mathrm{name}(v), \mathrm{def}(v) \rangle$ supplies the semantic context that constrains the proposed children to genuine refinements of $v$.

\noindent\textbf{Global Canonicalization} (Lines~\ref{ln:retrieve}--\ref{ln:add-edge}).
Each proposed child $\tilde{u} \in C(v)$ is treated as a topic \emph{mention} rather than inserted directly, because the same semantic topic is often discovered in different branches and at different recursion depths. \sys{} therefore resolves every mention against the global registry, the collection of all canonical topic nodes discovered so far, rather than only against siblings under the same parent or nodes at the same depth. Since comparing a mention against the entire registry is expensive, \sys{} first retrieves a compact candidate set $\mathcal{R}(\tilde{u})$ by embedding similarity over topic names and descriptions, and then asks the LLM to identify the candidates equivalent to the mention, producing $\mathcal{M}_{\mathrm{eq}}(\tilde{u})$. If an equivalent node exists, the mention is merged into it. Otherwise the mention becomes a new canonical node and is appended to the queue $\mathcal{Q}$ for further expansion. In both cases the registry holds only canonicalized topics rather than raw LLM output.
The canonicalized child is then linked to the current parent $v$. The edge direction follows from the top-down procedure, since children are proposed under the explicit instruction that they refine the parent. When the same canonical node is generated under different parents, it accumulates multiple incoming edges, naturally yielding a multi-parent DAG.

\noindent\textbf{Hierarchical Multi-Label Routing} (Lines~\ref{ln:route}--\ref{ln:update}).
After a node's children are canonicalized, \sys{} reroutes the papers assigned to the parent $v$ into its canonical children $C(v)$. We formulate routing as independent binary classification, one decision per child, rather than exclusive multiclass classification, because child topics may overlap and a paper may exhibit several distinct semantic facets. Concretely, for each paper $p \in D(v)$ and each child $u \in C(v)$, we prompt the LLM to decide whether $p$ belongs under $u$:
$
y_{p,u} = \mathrm{LLM}_{\text{route}}\big(\langle \mathrm{title}(p), \mathrm{abs}(p) \rangle,\, \langle \mathrm{name}(u), \mathrm{def}(u) \rangle\big).
$
The LLM takes the paper's title and abstract together with the child's name and definition, and outputs $y_{p,u} = 1$ if $p$ is routed to $u$ and $0$ otherwise. A paper may therefore be assigned to zero, one, or multiple children, which is consistent with the multi-parent DAG structure.

\noindent\textbf{Termination} (Lines~\ref{ln:terminate} and~\ref{ln:frontier}).
We stop expanding a node when either of two conditions holds: the number of papers routed to the node falls below a minimum size $\theta_{\min}$, in which case further splitting would produce topics with insufficient corpus support, or the node already lies at the maximum depth $L_{\max}$ of the DAG. The overall process terminates once no node in the queue $\mathcal{Q}$ remains expandable under these conditions. Finally, \sys{} extracts the frontier of each paper,
$
\Front{(A(p),\mathcal{G})}=\bigl\{v\in A(p):\nexists\,u\in A(p)\setminus\{v\}\ \text{s.t.}\ v\in\operatorname{Anc}^{+}(u)\bigr\},
$
which discards redundant ancestors while retaining multiple incomparable topics. 

%These frontiers form the paper representation used by the taxonomy-based similarity metric in Sec.~\ref{sec:optimization-framework}.

%% file: algs/taxonomy_construction.tex
\begin{algorithm}[t]
\footnotesize
\caption{Taxonomy construction}
\label{alg:taxonomy-construction}
\begin{algorithmic}[1]
\Require Accepted papers $\Papers$, root topic $r$, minimum node size $\theta_{\min}$, and maximum depth $L_{\max}$
\Ensure Topic DAG $\mathcal{G}=(\mathcal{V},\mathcal{E},r)$ and paper frontiers $\{F(p)\}_{p\in\Papers}$
\State Initialize $\mathcal{V}\gets\{r\}$, $\mathcal{E}\gets\emptyset$, $A(p)\gets\{r\}$ for all $p\in\Papers$, and $\mathcal{Q}\gets\{r\}$\label{ln:init}
\While{$\mathcal{Q}\neq\emptyset$}
    \State Pop $v$ from $\mathcal{Q}$ and set $D(v)\gets\{p\in\Papers:v\in A(p)\}$\label{ln:pop}
    \If{$|D(v)|<\theta_{\min}$ \textbf{ or } $\operatorname{depth}(v)\ge L_{\max}$}\label{ln:terminate}
        \State Mark $v$ as terminal and \textbf{continue}
    \EndIf
    \State $C(v)\gets\Call{ProposeChildrenLLM}{v,D(v)}$\label{ln:propose}
    \ForAll{$\tilde{u}\in C(v)$}
        \State $\mathcal{R}(\tilde{u})\gets\Call{Retrieve}{\tilde{u},\mathcal{V}}$\label{ln:retrieve}
        \State $\mathcal{M}_{\mathrm{eq}}(\tilde{u})\gets\Call{EquivalentLLM}{\tilde{u},\mathcal{R}(\tilde{u})}$\label{ln:eq}
        \State $u\gets\Call{Canonicalize}{\tilde{u},\mathcal{M}_{\mathrm{eq}}(\tilde{u}),\mathcal{V}}$\label{ln:canon}
        \State Update metadata of $u$, add $(v,u)$ to $\mathcal{E}$ if $u\neq v$, and enqueue $u$ into $\mathcal{Q}$ if $u$ is newly created\label{ln:add-edge}
    \EndFor
    \ForAll{$p\in D(v)$}
        \State $B(p)\gets\Call{RouteLLM}{p,\operatorname{Child}(v)}$\label{ln:route}
        \State $A(p)\gets A(p)\cup\bigcup_{u\in B(p)}\operatorname{Anc}^{+}(u)$\label{ln:update}
    \EndFor
\EndWhile
\State Set $F(p)\gets\operatorname{Frontier}(A(p),\mathcal{G})$ for all $p\in\Papers$\label{ln:frontier}
\State \Return $\mathcal{G}$ and $\{F(p)\}_{p\in\Papers}$
\end{algorithmic}
\end{algorithm}

%% file: sections/sec_4_solution_part_2.tex
\section{Constrained Taxonomy-Coherent Session Assignment}
\label{sec:optimization-framework}

The constructed taxonomy provides the semantic substrate for session organization. We cast the task as \noindent\textit{constrained taxonomy-coherent session assignment}: papers are partitioned into sessions so that papers sharing fine-grained taxonomy topics are grouped together, subject to the scheduling constraints of Sec.\ref{sec:problem}. To achieve this goal, we propose a new taxonomy-driven paper-to-paper distance that converts the topical structure of the taxonomy into a semantic cost between any pair of papers (Sec.\ref{sec:taxonomy-based-distance}). We then integrate this distance into a binary integer programming formulation whose objective rewards within-session topical coherence while enforcing the scheduling constraints, and describe how it be solved (Sec.\ref{sec:session-assignment-optimization}).

\subsection{Specificity-based Taxonomy Distance}
\label{sec:taxonomy-based-distance}

A natural distance metric on a hierarchy is the lowest-common-ancestor (LCA) distance, which compares two nodes by the depth of their deepest shared ancestor. Two properties of our taxonomy make this measure unsuitable. First, \sys{} refines different branches to different depths (Sec.\ref{sec:taxonomy-construction}), so edges are not semantically uniform. Two nodes at equal edge distance from the root may differ substantially in specificity, and any measure based on raw depth or path length conflates structural position with semantic content. Second, the taxonomy is a DAG rather than a tree. A multi-parent topic may have several incomparable common ancestors with another node, so the LCA itself is not well defined. We therefore construct a distance that (i) assigns each edge a length equal to the specificity gained along it, so that distance reflects topic specificity rather than depth, and (ii) replaces the LCA path with a shortest undirected path, which is well defined on a DAG and recovers the LCA form on trees. The measure is defined at two levels, a topic-level distance between taxonomy nodes and a paper-level distance between papers, with the former serving as the ground metric for the latter.

\noindent\textbf{Topic-level distance.}
Let $N=|\Papers|$, and let $n_v=|\{p\in\Papers:v\in A(p)\}|$ denote the support of topic $v$, that is, the number of accepted papers assigned to it. Because unsupported topics are removed during construction, every node has positive support. The empirical probability of topic $v$ is $\Pr(v)=n_v/N$, and its \emph{specificity} is its empirical information content~\cite{resnik1995ic,lin1998similarity}, $\operatorname{IC}(v)=-\log\Pr(v)$, analogous to an inverse document frequency over the accepted papers. The root satisfies $n_r=N$ and therefore $\operatorname{IC}(r)=0$. Upward closure implies that a parent has support at least as large as any of its descendants, and construction enforces $n_v<n_u$ on every accepted edge $u\rightarrow v$. Consequently, $\operatorname{IC}(v)>\operatorname{IC}(u)$, so every accepted specialization edge receives a positive length.

For each directed specialization edge $u\rightarrow v$, with $u$ broader and $v$ narrower, the edge length is the \emph{specificity gain} $\ell(u,v)=\operatorname{IC}(v)-\operatorname{IC}(u)$; by upward closure it equals $-\log\Pr(v\mid u)$, the conditional surprisal of specializing $u$ to $v$. Let $\overline{\mathcal{G}}$ denote the undirected taxonomy obtained by dropping edge directions while retaining these lengths. The topic distance between nodes $u$ and $v$ is the shortest undirected path length
\begin{equation}
    \delta_{\mathcal{G}}(u,v) = \min_{\pi:u\leadsto v} \sum_{(a,b)\in\pi} \ell(a,b),
    \label{eq:topic-distance}
\end{equation}
where $\pi$ ranges over all undirected paths between $u$ and $v$ in $\overline{\mathcal{G}}$. Because the taxonomy is rooted, every node is connected to the root, so such a path always exists. When $\mathcal{G}$ is a tree, Equation~\eqref{eq:topic-distance} reduces to the familiar form $\delta_{\mathcal{G}}(u,v)=\operatorname{IC}(u)+\operatorname{IC}(v)-2\operatorname{IC}(\operatorname{LCA}(u,v))$. On a DAG, the shortest-path definition generalizes this form to multi-parent topics, where two nodes may be connected through more than one meaningful ancestor path.

\noindent\textbf{Paper-level distance.}
A paper may be associated with several frontier topics, so \sys{} represents each paper not by a single node but by the uniform distribution over its frontier, $\mu_p=|F(p)|^{-1}\sum_{u\in F(p)}\Delta_u$, where $\Delta_u$ is a point mass at topic $u$. The taxonomy-based distance between papers $p$ and $q$ is the optimal-transport distance between their frontier distributions under the ground distance $\delta_{\mathcal{G}}$:
\begin{equation}
    d^{\mathrm{tax}}_{pq}
    =
    \min_{\Gamma\ge 0}
    \sum_{u\in F(p)} \sum_{v\in F(q)}
    \Gamma_{uv}\delta_{\mathcal{G}}(u,v),
    \label{eq:paper-ot-distance}
\end{equation}
subject to $\sum_{v\in F(q)}\Gamma_{uv}=1/|F(p)|$ for all $u\in F(p)$ and $\sum_{u\in F(p)}\Gamma_{uv}=1/|F(q)|$ for all $v\in F(q)$. This compares two papers by the minimum semantic cost of transporting one paper's topic mass onto the other's. The resulting $d^{\mathrm{tax}}_{pq}$ is the 1-Wasserstein distance between their uniform frontier distributions and constitutes a valid metric over their taxonomy-based representations. %it is nonnegative and symmetric, satisfies the triangle inequality, and equals zero exactly when the two frontier distributions coincide, as proved in Appendix~\ref{app:optimization-proofs}.

\subsection{Constraint-Aware Optimization Formulation}
\label{sec:session-assignment-optimization}
We now incorporate the specificity-based taxonomy distance into the \taskname{} task of Sec.\ref{sec:problem} by formulating it as a binary integer linear programming problem. The formulation consists of a coherence objective built on the precomputed distances and a set of hard constraints that encode the scheduling requirements.

\noindent\textbf{Objective: taxonomy-based coherence.}
The model uses three families of binary variables. Variable $x_{ps}$ indicates whether paper $p$ is assigned to session $s$, $y_s$ indicates whether session $s$ is activated, and $z_{pqs}$ (for $p<q$) indicates whether papers $p$ and $q$ share session $s$. Their domains are fixed by Equation~\eqref{eq:binary-domain}. The taxonomy enters the formulation only through the precomputed pairwise distances $d_{pq}=d^{\mathrm{tax}}_{pq}$ of Sec.\ref{sec:taxonomy-based-distance}, which act as constant coefficients. The objective minimizes the total within-session semantic distance:
\begin{equation}
    \min \sum_{s\in\Sessions}
    \sum_{\substack{p,q\in\Papers}}
    \omega_s d_{pq} z_{pqs}.
    \label{eq:session-objective}
\end{equation}
Here $\omega_s=2/(c_s(c_s-1))$ is the session weight, which normalizes each session's contribution by its number of paper pairs so that sessions of different capacities are weighted comparably. Because the objective depends on whether two papers share a session, which is the product $x_{ps}x_{qs}$, we tie $z_{pqs}$ to this product through the standard linearization
\begin{equation}
    z_{pqs}\le x_{ps},
    \;\;
    z_{pqs}\le x_{qs},
    \;\;
    z_{pqs}\ge x_{ps}+x_{qs}-1,
    \quad \forall p<q,\ \forall s\in\Sessions.
    \label{eq:z-linear}
\end{equation}
This construction cleanly separates semantic modeling from combinatorial assignment: all taxonomy reasoning is confined to the precomputation of $d_{pq}$, and the assignment formulation itself is a binary integer linear programming problem with a linear objective.

\noindent\textbf{Linear constraints.}
We next show how the scheduling requirements of Sec.\ref{sec:problem} are encoded as linear constraints.

\noindent\textit{\underline{Assignment completeness.}}
Each paper is assigned to exactly one session:
\begin{equation}
    \sum_{s\in\Sessions}x_{ps}=1,
    \qquad \forall p\in\Papers.
    \label{eq:assign-once}
\end{equation}

\noindent\textit{\underline{Session capacity and activation.}}
An activated session holds at least $\ell_s$ and at most $c_s$ papers, while an inactive session receives none:
\begin{equation}
    \ell_s y_s \le \sum_{p\in\Papers}x_{ps} \le c_s y_s,
    \qquad \forall s\in\Sessions.
    \label{eq:capacity}
\end{equation}
If all candidate sessions are mandatory, we fix $y_s=1$ for every $s\in\Sessions$. If exactly $K$ sessions should be activated, we add $\sum_{s\in\Sessions}y_s=K$.

\noindent\textit{\underline{Track and placement compatibility.}}
A paper may be assigned only to a session that permits its track and that lies in its allowed set, where $\mathcal{S}^{\mathrm{place}}_p$ denotes the sessions allowed for paper $p$:
\begin{equation}
    x_{ps}=0,
    \qquad
    \forall p\in\Papers,\ \forall s\ \text{with}\ t_p\notin U_s\ \text{or}\ s\notin\mathcal{S}^{\mathrm{place}}_p.
    \label{eq:track-placement}
\end{equation}

\noindent\textit{\underline{Presenter conflicts.}}
Let $\mathcal{E}_{\mathrm{conf}}$ be the set of paper pairs that share a presenter and let $\Sessions_k=\{s\in\Sessions:k_s=k\}$ be the sessions in time slot $k$. A conflicting pair cannot occupy two different sessions in parallel:
\begin{equation}
    x_{ps}+x_{qt}\le 1
    \ \ \text{and}\ \
    x_{qs}+x_{pt}\le 1,
    \quad
    \forall (p,q)\in\mathcal{E}_{\mathrm{conf}},
    \ \forall s,t\in\Sessions_k,\ s\ne t.
    \label{eq:presenter-conflict}
\end{equation}
These constraints permit the two papers to share a single session but forbid scheduling them in parallel sessions within the same slot.

\noindent\textit{\underline{Binary domains.}}
In addition to the scheduling requirements above, all decision variables are binary:
\begin{equation}
    x_{ps},y_s,z_{pqs}\in\{0,1\}.
    \label{eq:binary-domain}
\end{equation}

\noindent\textbf{Exact and Fallback Solution.}
The formulation given by Objective~\eqref{eq:session-objective} and Constraints~\eqref{eq:z-linear}--\eqref{eq:binary-domain} is a mixed-integer linear programming problem with binary variables, and we solve it exactly with branch-and-bound and branch-and-cut methods~\cite{wolsey2020integer} as implemented in off-the-shelf solvers such as Gurobi~\cite{gurobi}, SCIP~\cite{scip}, and OR-Tools CP-SAT~\cite{ortools}, which return an optimal assignment whenever the constraint system is feasible. When the constraints are too harsh to admit any feasible solution, \sys{} falls back to a soft-constraint reformulation: selected hard constraints are converted into penalty terms in the objective, weighted according to the organizers' preferences over the constraints, so that the reformulation remains a mixed-integer linear programming problem and is solved with the same solvers. We provide the detailed reformulation and a time-complexity analysis in Appendix~\ref{app:solving}, and empirically show in Sec.~\ref{sec:efficiency} that the solving time is acceptable in practice.

%% file: sections/sec_5_solution_part_3.tex
\section{Final Program Refinement}
\label{sec:post-organization-finalization}

Given the grouped papers as sessions, \sys{} further refines and adjusts the program to accommodate other soft preferences that make it more usable for attendees, without changing the paper grouping within any session.

\noindent\textbf{Session title generation.}
For each session $s$ with papers $\Papers_s=\{p:x_{ps}=1\}$, \sys{} generates a title by prompting an LLM with two inputs per paper: the paper information (title and abstract) and its taxonomy path information, namely the root-to-frontier paths of its assigned topics in $\mathcal{G}$. The paths situate each paper within the hierarchy, so the LLM observes not only what the papers are about but also at which level of specificity their topics agree. The LLM is instructed to produce a short title (three to eight words) that captures the shared themes rather than concatenating keywords.

\noindent\textbf{Time-slot alignment.}
Because attendees can attend only one of several parallel sessions, co-scheduling topically related sessions forces them to choose between overlapping content and miss talks they would otherwise attend. \sys{} therefore assigns sessions to time slots so that topically divergent sessions run in parallel, sparing attendees from clashes between closely related sessions. To compare sessions, \sys{} represents each session $s$ by an embedding $e_s$ that averages the encoding of its generated title with the encodings of its papers,
$
e_s=\frac{1}{1+|\Papers_s|}\Bigl(\phi(\mathrm{title}_s)+\sum_{p\in\Papers_s}\phi(p)\Bigr),
$
where $\phi(\cdot)$ is a text encoder applied to the session title and to each paper's title and abstract. The divergence between sessions $s$ and $t$ is their cosine dissimilarity,
$
    D_{st}=1-\frac{\langle e_s,e_t\rangle}{\lVert e_s\rVert\,\lVert e_t\rVert}.
    \label{eq:session-divergence}
$
Let $a_{sk}\in\{0,1\}$ indicate whether session $s$ is placed in slot $k\in\mathcal{K}$. The alignment maximizes the total divergence among co-scheduled sessions,
$
    \max
    \sum_{k\in\mathcal{K}}
    \sum_{\substack{s,t\in\Sessions\\s<t}}
    D_{st} b_{stk},
    \label{eq:slot-divergence-objective}
$
with $b_{stk}$ linearized by $b_{stk}\le a_{sk}$, $b_{stk}\le a_{tk}$, and $b_{stk}\ge a_{sk}+a_{tk}-1$, subject to one slot per session, $\sum_{k}a_{sk}=1$, and a per-slot cap $\sum_{s}a_{sk}\le M_k$ set by the room count or the desired parallelism. Fixed sessions, unavailable speakers, and track-specific restrictions are imposed by fixing the corresponding $a_{sk}$. As in the session-assignment stage (Section~\ref{sec:optimization-framework}), this is a binary integer linear programming problem and is solved with the same solver. Because it ranges over sessions rather than papers, the problem is small and is solved to optimality directly.

\noindent\textbf{Popularity prediction for room alignment.}
Matching room capacity to session demand matters for attendees: a popular session placed in a small room leaves interested attendees unable to get a seat, while an over-provisioned room wastes capacity that another session could have used. \sys{} therefore places higher-demand sessions in larger rooms. Since demand is unknown before the conference, \sys{} estimates the popularity of each session from the past $N$ editions of the conference. For a paper $p$ and a historical paper $h\in\mathcal{H}_N$, let $\operatorname{sim}(p,h)=\langle\phi(p),\phi(h)\rangle/(\lVert\phi(p)\rVert\,\lVert\phi(h)\rVert)$ be their cosine embedding similarity, and let $\mathcal{H}_p=\{h\in\mathcal{H}_N:\operatorname{sim}(p,h)\ge\tau\}$ collect the historical papers related to $p$ under threshold $\tau$. The popularity of session $s$ sums the relevance-weighted citations of the historical papers related to its papers,
$
\rho_s=\log\Bigl(1+\sum_{p\in\Papers_s}\sum_{h\in\mathcal{H}_p}\operatorname{sim}(p,h)\,\operatorname{cit}(h)\Bigr),
$
with the logarithm damping citation outliers. Room alignment then runs independently within each time slot. \sys{} ranks the sessions scheduled in the slot by popularity $\rho_s$ and the available rooms by capacity $C_r$, and pairs them in rank order, so the most popular session takes the largest room. 

\noindent\textbf{Board placement for the poster program.} Poster sessions raise a within-session problem that oral scheduling does not: each accepted poster must be placed on a physical board, and attendees browse the boards by walking past them in order, so related posters should sit on nearby boards. \sys{} accounts for two factors, presenter proximity and topical proximity, and reformulates board placement as a one-dimensional sequencing problem: common physical layouts induce a natural linear traversal of the boards (a single row is already a sequence, a grid is traversed in serpentine order, and a cyclic loop is cut into a path), so placing posters along this traversal makes sequence adjacency approximate physical proximity on the floor. The full formulation and solution procedure are given in Appendix~\ref{app:board-placement}.

%% file: sections/sec_6_experiment.tex
\section{Experiment}
\label{sec:experiments}

\subsection{Experiment Setting}

\noindent\textbf{Benchmarks.} 
We build our benchmarks from the official websites of four top-tier computer science conferences, which publish, for each accepted paper, its title, author list, and the human-curated session to which organizers assigned it. These human-curated sessions serve as expert-constructed reference programs against which we evaluate automatically generated ones. Because the official listings do not include abstracts, we enrich each paper with its abstract retrieved from OpenAlex~\cite{priem2022openalex}, matching entries by title and authors.  Table~\ref{tab:dataset_statistics} summarizes the benchmark statistics. We focus on oral sessions, since official programs provide human-curated topical groupings only for oral papers, whereas poster placements carry no comparable reference partition. The poster pipeline is instead evaluated in our deployment study and attendee survey (Sec.~\ref{sec:deployment}).

\input{tables/dataset_statistics}

\noindent\textbf{Evaluation Metrics.}
We evaluate program quality from three complementary perspectives:
\begin{itemize}[noitemsep,leftmargin=*,topsep=0pt]
\item \textbf{LLM evaluation.}
This perspective measures intrinsic session quality without reference to a human program, so it can credit coherent groupings that differ from the committee's partition. \emph{Coherence} (Coh.): an LLM evaluator scores each session on a $1$--$5$ scale from its paper titles and abstracts, where higher scores indicate that the papers share a common problem, method, or application. \emph{Outliers} (Out.): the same evaluator flags papers weakly related to their session theme, reported as the number of outlier papers.
\item \textbf{Human agreement.}
This perspective measures how well the predicted sessions recover the human-curated reference program, treating session organization as a clustering of papers. We report three standard clustering-agreement metrics: normalized mutual information (NMI)~\citep{strehl2002cluster,vinh2010information}, the adjusted Rand index (ARI)~\citep{hubert1985comparing}, and cluster-level F1~\citep{manning2008introduction}. %NMI measures the information shared between the predicted and reference partitions, ARI corrects pair-counting agreement for chance, and cluster-F1 balances precision and recall over co-assigned paper pairs.
\item \textbf{Constraint violation.}
This perspective checks whether the program satisfies practical design constraints. \emph{Capacity} (Cap.): the number of sessions whose sizes fall outside the allowed range. \emph{Presenter conflicts} (Conf.): the number of presenter-sharing paper pairs scheduled in parallel sessions within the same time slot.
\end{itemize}

\noindent\textbf{Baselines.}
We compare \sys against two families of baselines: optimization-based methods and LLM-based methods.

\noindent\textit{\underline{Optimization-based methods.}}
These methods solve the same constrained assignment problem that \sys{} uses (Sec.~\ref{sec:optimization-framework}), assigning papers to sessions by minimizing Objective~\eqref{eq:session-objective} under the capacity, track, placement, and presenter-conflict constraints. They differ from \sys{} only in the pairwise cost that drives the objective: the taxonomy-based distance $d_{pq}$ is replaced by a distance derived from a flat pairwise similarity, so the comparison isolates the effect of the taxonomy-induced metric. The two variants differ in how that similarity is computed: 1) \emph{\TopicILP} represents each paper by a topic distribution inferred with a topic model~\cite{blei2003lda} and compares distributions in that space; 2) \emph{\EmbILP} represents each paper by a dense embedding from SPECTER2~\cite{singh2023scirepeval}, a language model widely adopted for scientific text, and uses cosine similarity.

\noindent\textit{\underline{LLM-based methods.}}
These methods prompt an LLM to organize the papers directly into the required number of sessions from their titles, abstracts, and keywords, under the same capacity and track constraints. We evaluate two variants: 1) \emph{Direct prompting} (\LLMDirect) produces the full session assignment in a single generation; 2) \emph{ReAct-style prompting} (\LLMReAct)~\cite{yao2023react,madaan2023selfrefine} first proposes an initial clustering, then checks the resulting sessions against semantic and constraint-based criteria, and revises the assignment accordingly.

\noindent\textbf{Implementation Configuration.}
All experiments are conducted on an Ubuntu server equipped with an Intel Xeon Platinum 8470Q processor and 90\,GB of memory. Full implementation details are provided in Appendix~\ref{sec:impl-details}. The datasets are released in~\citep{taxoconf_material}, and \sys{} is publicly available as a web application at \url{https://taxoconf.com}.

\input{tables/effectiveness}

\subsection{Offline Evaluation}\label{sec:main_results}

\subsubsection{Program Quality Evaluation.} Table~\ref{tab:main_results} reports the effectiveness of \sys{} on four retrospective benchmarks. \sys{} performs the best on all metrics across all benchmarks: it ranks first on coherence, outliers, NMI, ARI, and cluster-F1 with zero constraint violations, improving over the strongest baseline by 4\% on coherence, 12\% on NMI, and 33\% on cluster-F1 on average, and the gains are stable across venue types (10\%--14\% on NMI over ICML, ICLR, KDD, and ICSE). The two baseline families are limited in complementary ways. Optimization baselines are feasible but less aligned with human programs: \EmbILP{} beats \TopicILP{} on every metric yet still trails \sys{}, as flat embedding similarity misses the hierarchical, cross-cutting structure human organizers use. LLM baselines invert this profile: they are locally coherent (\LLMReAct{} within about 4\% of \sys{} on coherence) but structurally and operationally weak, trailing \EmbILP{} on NMI and incurring 96 and 58 violations without an optimization layer to enforce hard constraints. The improvements are attributed to the taxonomy-induced distance: the strongest baseline on the agreement metrics, \EmbILP{}, solves the identical assignment program under the same constraints, so \sys{}'s gains isolate the effect of replacing flat embedding similarity with the taxonomy-induced pairwise cost $d_{pq}$, which delivers both higher coherence and closer human agreement at no loss of feasibility. An ablation of each component is reported in Appendix~\ref{app:ablation}.

\subsubsection{Efficiency and Monetary Cost Comparison.}\label{sec:efficiency} 
\begin{figure}[t]
  \centering
  \includegraphics[width=\linewidth]{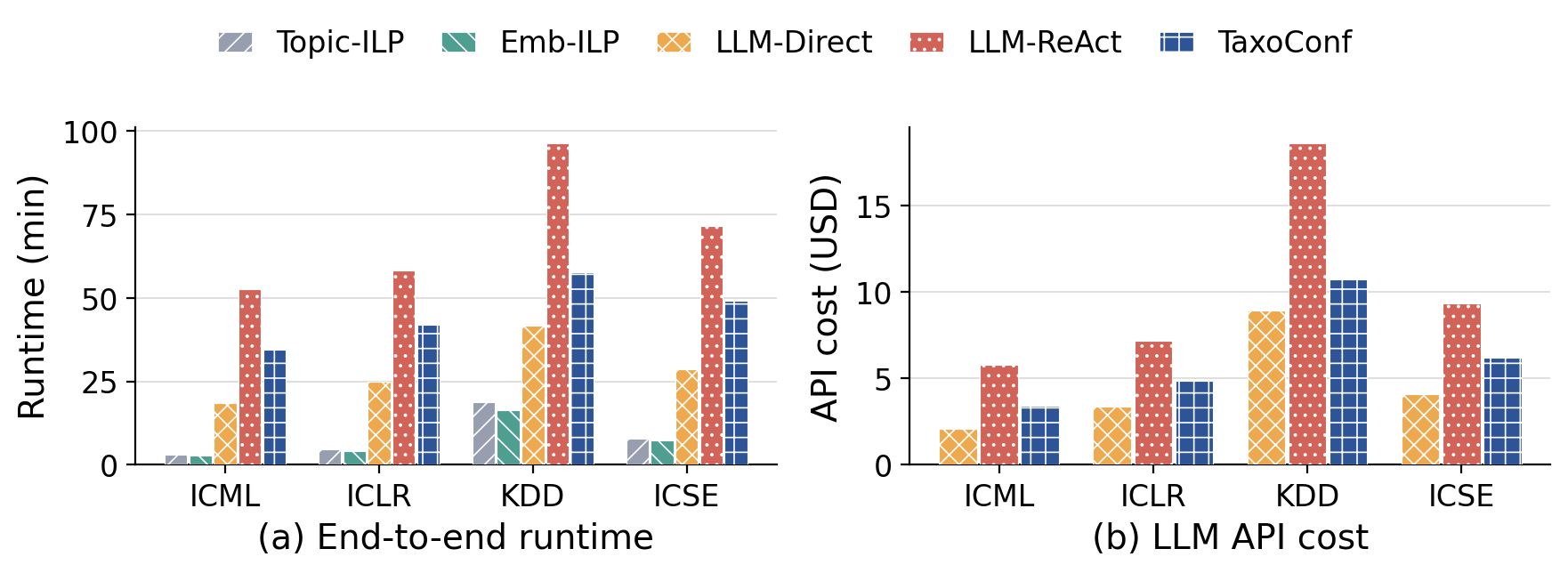}
  \caption{Efficiency and monetary cost across the benchmarks.}
  \label{fig:efficiency}
\end{figure}

Fig.~\ref{fig:efficiency} presents the runtime and monetary cost of each method. The optimization baselines (\TopicILP, \EmbILP) are the most efficient and incur no API cost, since they reuse precomputed representations and invoke no LLM queries, but their session quality is the weakest in Table~\ref{tab:main_results}. Among the LLM-based methods, \LLMReAct is the slowest and most expensive because its propose-check-revise loop issues requires repeated LLM calls over the full paper set across several rounds, whereas \LLMDirect generates one assignment in a single pass. \sys falls between the two: it invokes the LLM once to construct the taxonomy and then resolves assignment and placement with a solver, so it avoids the iterative re-reasoning of \LLMReAct while still using the LLM where it adds value. On average \sys runs in 45.9 minutes at \$6.30, a 34\% runtime reduction and a 38\% cost reduction relative to \LLMReAct, the strongest baseline in Table~\ref{tab:main_results}. \LLMDirect is cheaper still (\$4.61), but produces lower-quality, constraint-violating programs. Both the runtimes and the API costs are modest relative to the manual process, which organizers currently perform over a much longer period.

%% file: tables/dataset_statistics.tex
\begin{table}[t]
\centering
\small
\caption{Statistics of the created benchmarks.}
\label{tab:dataset_statistics}
\begin{tabular}{lrrr}
\toprule
Dataset & Sessions & Papers & Session Capacity \\
\midrule
ICML25 & 30 & 120 & 4 \\
ICLR25 & 36 & 213 & 5--6 \\
KDD25 & 132 & 771 & 4--6 \\
ICSE25 & 65 & 242 & 3--6 \\
\bottomrule
\end{tabular}
\end{table}

%% file: tables/effectiveness.tex
\begin{table}[t]
\centering
\scriptsize
\setlength{\tabcolsep}{2.5pt}
\renewcommand{\arraystretch}{1.0}
\caption{Overall per-conference results on the four benchmarks. Arrows give the preferred direction ($\uparrow$ higher is better, $\downarrow$ lower is better). The best result is in \textbf{bold}.}
\label{tab:main_results}
\begin{tabular}{@{}cl cc ccc cc@{}}
\toprule
\multirow{2}{*}{\textbf{Venue}} & \multirow{2}{*}{\textbf{Method}} & \multicolumn{2}{c}{\textbf{LLM Eval.}} & \multicolumn{3}{c}{\textbf{Human Agreement}} & \multicolumn{2}{c}{\textbf{Violation}} \\
\cmidrule(lr){3-4}\cmidrule(lr){5-7}\cmidrule(lr){8-9}
& & \textbf{Coh.}$\uparrow$ & \textbf{Out.}$\downarrow$ & \textbf{NMI}$\uparrow$ & \textbf{ARI}$\uparrow$ & \textbf{F1}$\uparrow$ & \textbf{Cap.}$\downarrow$ & \textbf{Conf.}$\downarrow$ \\
\midrule
\multirow{5}{*}{\rotatebox[origin=c]{90}{ICML25}} & \TopicILP & 4.11 & 24 & 0.583 & 0.082 & 0.352 & \textbf{0} & \textbf{0} \\
& \EmbILP & 4.15 & 20 & 0.620 & 0.112 & 0.397 & \textbf{0} & \textbf{0} \\
& \LLMDirect & 4.31 & 13 & 0.662 & 0.105 & 0.400 & 5 & 3 \\
& \LLMReAct & 4.45 & 10 & 0.672 & 0.122 & 0.433 & 3 & 2 \\
\cmidrule(lr){2-9}
& \textbf{\sys} & \textbf{4.64} & \textbf{5} & \textbf{0.748} & \textbf{0.219} & \textbf{0.541} & \textbf{0} & \textbf{0} \\
\midrule
\multirow{5}{*}{\rotatebox[origin=c]{90}{ICLR25}} & \TopicILP & 4.09 & 37 & 0.571 & 0.105 & 0.382 & \textbf{0} & \textbf{0} \\
& \EmbILP & 4.13 & 33 & 0.628 & 0.149 & 0.436 & \textbf{0} & \textbf{0} \\
& \LLMDirect & 4.34 & 22 & 0.662 & 0.204 & 0.468 & 7 & 4 \\
& \LLMReAct & 4.46 & 17 & 0.665 & 0.208 & 0.472 & 4 & 3 \\
\cmidrule(lr){2-9}
& \textbf{\sys} & \textbf{4.65} & \textbf{9} & \textbf{0.759} & \textbf{0.286} & \textbf{0.586} & \textbf{0} & \textbf{0} \\
\midrule
\multirow{5}{*}{\rotatebox[origin=c]{90}{KDD25}} & \TopicILP & 4.12 & 112 & 0.585 & 0.080 & 0.298 & \textbf{0} & \textbf{0} \\
& \EmbILP & 4.17 & 98 & 0.607 & 0.104 & 0.333 & \textbf{0} & \textbf{0} \\
& \LLMDirect & 4.29 & 66 & 0.683 & 0.071 & 0.296 & 38 & 24 \\
& \LLMReAct & 4.39 & 52 & 0.696 & 0.098 & 0.331 & 22 & 14 \\
\cmidrule(lr){2-9}
& \textbf{\sys} & \textbf{4.61} & \textbf{27} & \textbf{0.768} & \textbf{0.174} & \textbf{0.471} & \textbf{0} & \textbf{0} \\
\midrule
\multirow{5}{*}{\rotatebox[origin=c]{90}{ICSE25}} & \TopicILP & 4.03 & 47 & 0.546 & 0.098 & 0.352 & \textbf{0} & \textbf{0} \\
& \EmbILP & 4.09 & 40 & 0.616 & 0.147 & 0.386 & \textbf{0} & \textbf{0} \\
& \LLMDirect & 4.36 & 25 & 0.681 & 0.062 & 0.380 & 9 & 6 \\
& \LLMReAct & 4.41 & 21 & 0.689 & 0.075 & 0.396 & 6 & 4 \\
\cmidrule(lr){2-9}
& \textbf{\sys} & \textbf{4.58} & \textbf{10} & \textbf{0.767} & \textbf{0.289} & \textbf{0.566} & \textbf{0} & \textbf{0} \\
\bottomrule
\end{tabular}
\end{table}

%% file: sections/sec_7_deployment.tex
\subsection{Deployment and Post-Launch Performance}
\label{sec:deployment}

\sys{} has been deployed in real-world conference-organization workflows, including IJCNN~2026\footnote{Held as part of WCCI~2026: \url{https://attend.ieee.org/wcci-2026/daily-program/}.} and SIGIR~2026\footnote{\url{https://sigir2026.org/program.html}}. It is also slated for adoption by upcoming conferences, including VLDB~2026, SIGIR~2027, ICDE~2027, and DASFAA~2027. In each deployment, \sys{} generated an initial program that the organizers then inspected, revised, and approved before publication. In this section, we quantify how much organizer intervention was required to transform the generated program into the published one. We center the analysis on SIGIR~2026, whose program involves richer scheduling constraints than those of IJCNN~2026.

Beyond topical coherence, the organizers specified a set of hard operational constraints that any usable program must satisfy. We group these constraints by session type. For oral sessions, \textbf{(i)~Presenter conflicts:} because the schedule is fixed before registration closes, when the eventual presenter of each paper is still unknown, papers whose author sets overlap must not be placed in parallel sessions. \textbf{(ii)~Session composition:} each oral session admits at most three non-full papers, including short, reproducibility, perspective, resource, and low-resource submissions, with the sole exception of one dedicated Low-Resource session pinned to a fixed slot. \textbf{(iii)~Panelist availability:} the three named industry-panel discussants must remain free during the panel, so none of their papers is scheduled opposite it. \textbf{(iv)~Room allocation:} rooms are assigned according to predicted session popularity, so that higher-demand sessions receive larger rooms. For poster sessions, \textbf{(v)~Powered demonstrations:} demonstrations are placed only on the twelve powered poster boards. \textbf{(vi)~Author spread:} posters sharing an author are distributed across the two poster days. \textbf{(vii)~Neighborhood placement:} within each poster session, posters that are topically similar or share authors are placed on nearby boards.

\subsubsection{Post-Launch Performance.} The output covered all 657 accepted papers, comprising 293 oral presentations assigned to 51 topical oral sessions and 364 posters assigned to three poster sessions in the 132-board exhibition hall. 

\noindent\textbf{Organizer post-editing.} Table~\ref{tab:sigir_postlaunch} summarizes how much of the generated program was preserved through organizer review. The initial \sys{} output was already operationally feasible, with zero hard-constraint violations requiring organizer correction across both the oral and poster programs. Beyond feasibility, the initial and final schedules agree closely, with 91.8\% of oral assignments preserved and only 8.2\% of papers moved during revision. This corresponds to 0.47 moved papers per session and an average session edit distance of 0.94 paper edits, indicating that the final oral program was obtained through localized refinements rather than substantial reconstruction. Only two of the 51 topical sessions required regrouping, and only three session titles were edited. The poster program required no post-editing at all: all 364 posters were deployed on their generated boards, with no reassignment across sessions or boards.

\begin{figure}
    \centering
    \includegraphics[width=0.9\linewidth]{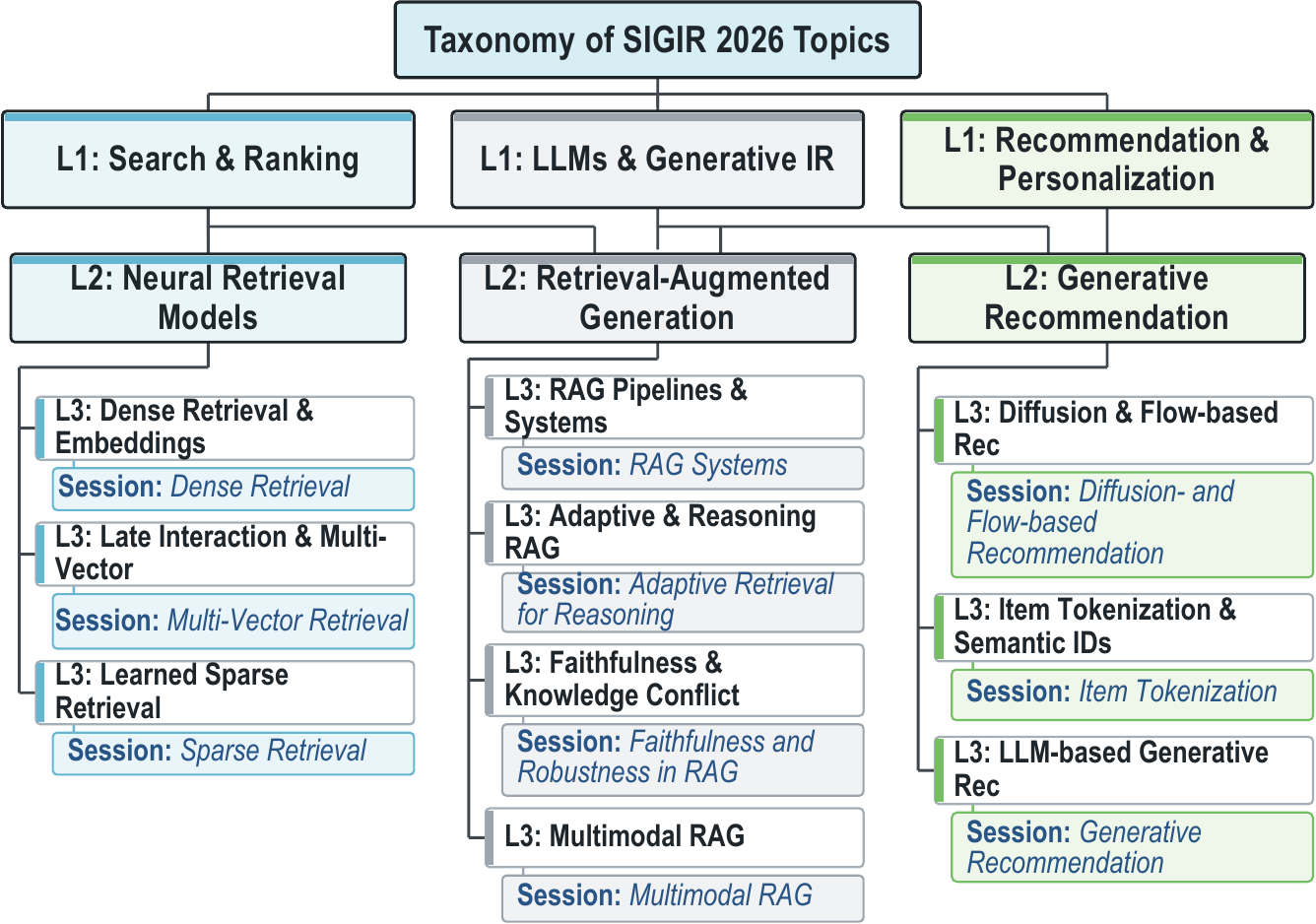}
    \caption{An excerpt of the generated taxonomy and the oral sessions derived from its frontier topics.}
    \label{fig:case_study}
\end{figure}

\noindent\textbf{Attendee survey.}
To assess how the deployed program was received, we conducted an anonymous on-site survey of SIGIR~2026 participants, covering ten aspects of the program on a 5-point scale. The program was rated favorably: 46 of the 47 respondents rated the overall program quality 4 or higher (mean 4.26), and the aspects that \sys{} directly controls, including oral session coherence (4.21), session title accuracy and wording (both 4.15), parallel-session separation (4.09), and poster placement (4.06), all received a median of 4. The lowest-rated aspect was the adequacy of room sizes (3.77), and crowded rooms for popular sessions were also the most frequent free-text complaint; consistent with this, the general chair reported that only one of the 51 oral sessions, \emph{LLM-based Evaluation and Relevance Assessment}, drew an audience that exceeded its room capacity, likely because the citation-based popularity estimate fails to capture the rapidly rising interest in emerging topics whose papers have not yet accumulated citations. This points to popularity-based room allocation as the primary refinement target for future work.  Appendix~\ref{sec:attendee-survey} reports the respondent distribution and per-aspect results.

\noindent\textbf{A visual explanation: from taxonomy to schedule.}
The published program shows how the taxonomy structure carries through to the deployed schedule; Fig.~\ref{fig:case_study} illustrates a local excerpt of the taxonomy and the sessions it induced. The 51 oral session themes sit at the granularity of the taxonomy frontier rather than at coarse area labels: recommendation is resolved into more than a dozen sessions distinguished by technique and setting (e.g., \emph{Sequential Recommendation}, \emph{Diffusion- and Flow-based Recommendation}, \emph{Item Tokenization}, \emph{CTR Prediction}), and retrieval-augmented generation into distinct facets (\emph{RAG Systems}, \emph{Faithfulness and Robustness in RAG}, \emph{Multimodal RAG}, \emph{Adaptive Retrieval for Reasoning}). Individual sessions are coherent at the method-family level: \emph{Multi-Vector Retrieval} gathers six late-interaction papers---index compression, two token-pruning studies, numeracy injection, and pseudo-relevance feedback for PLAID---including two reproducibility papers, showing that the taxonomy groups papers across submission tracks and heterogeneous title wording. Conversely, the taxonomy keeps apart what surface keywords would conflate: papers in both \emph{Item Tokenization} and \emph{Generative Recommendation} almost uniformly mention generative recommendation in their titles, yet the former collects semantic-ID and tokenization work while the latter collects generation-side training and preference-optimization work, mirroring two distinct frontier topics. Time-slot alignment placed sibling sessions in disjoint slots: for example, \emph{Dense}, \emph{Sparse}, and \emph{Multi-Vector Retrieval} run on different days or slots, so attendees following one line of work face no forced choice between closely related sessions.

\input{tables/post-launching_performance}

%% file: tables/post-launching_performance.tex
\begin{table}[t]
\centering
\caption{Evaluation for SIGIR~2026 Program Quality.}
\label{tab:sigir_postlaunch}
\small
\begin{tabular}{@{}lr@{}}
\toprule
\textbf{Metric} & \textbf{Value} \\
% \midrule
% \multicolumn{2}{@{}l}{\textit{Program scale}} \\
% Accepted papers & 657 \\
% Oral presentations & 293 \\
% Poster presentations & 364 \\
% Oral sessions & 51 \\
% Poster sessions & 3 \\
% Poster boards & 132 \\
\midrule
\multicolumn{2}{@{}l}{\textit{Oral-program preservation}} \\
Assignment preservation & 91.8\% (269/293) \\
Paper movement & 8.2\% (24/293) \\
Per-session paper movement & 0.47 papers/session \\
Sessions needing regrouping & 2/51 (3.9\%) \\
Average session edit distance & 0.94 paper edits/session \\
Session title edits & 3/51 (5.9\%) \\
\midrule
\multicolumn{2}{@{}l}{\textit{Poster-program preservation}} \\
Poster reassignments & 0/364 (0\%) \\
\midrule
\multicolumn{2}{@{}l}{\textit{Constraint satisfaction (oral and poster)}} \\
Hard-constraint violations corrected & 0 \\
\bottomrule
\end{tabular}
\end{table}

%% file: sections/sec_8_related_work.tex
\section{Related Work}
\label{sec:related-work}
\noindent\textbf{\TaskName{}.}
Existing approaches fall into three families, each trading one requirement for another. Human-in-the-loop systems elicit affinity judgments or metadata from the community to support interactive session construction~\cite{kim2013cobi,andre2013community,chilton2014frenzy}; although effective, they demand substantial additional input from participants and organizers, which limits their scalability. Optimization-based methods formulate session assignment and conference scheduling as constrained optimization problems~\cite{sidiropoulos2015signal,bulhoes2022conference}; they enforce operational constraints explicitly, but measure topical relatedness by flat pairwise similarities from topic models or text embeddings, which overlook the hierarchical and cross-cutting structure among research topics. LLM-based approaches prompt an LLM to construct the program directly or incorporate LLM-derived similarities into an integer-programming formulation~\cite{jobson2024investigating}; they provide richer semantic organization but do not jointly enforce conference-wide operational requirements. \sys{} resolves these trade-offs without crowd effort: it derives session composition, session titles, and the paper-to-paper distance from an explicitly constructed hierarchical taxonomy, and couples the taxonomy-based distance with a constraint-aware integer programming formulation, achieving topical coherence and operational constraints simultaneously.

\noindent\textbf{Taxonomy Construction.}
Automatic taxonomy construction organizes terms, entities, topics, or documents into hierarchical structures. Existing methods can be grouped into rule-based, learning-based, and LLM-based approaches. Rule-based methods extract taxonomic relations using manually designed lexical or syntactic patterns, such as Hearst patterns for hyponym acquisition~\cite{hearst1992automatic} and semi-supervised web-based bootstrapping from seed concepts~\cite{kozareva2010semi}. Learning-based methods infer taxonomic structure from corpus statistics, term embeddings, graph structure, or self-supervised signals, supporting both taxonomy induction from scratch and the expansion of existing taxonomies~\cite{snow2006semantic,zhang2018taxogen,shen2018hiexpan,shang2020nettaxo,shen2020taxoexpan,huang2020corel}. More recently, LLM-based methods use large language models to generate, refine, or adapt taxonomies, including label taxonomy generation~\cite{wan2024tntllm}, layer-by-layer taxonomy induction~\cite{zeng2024chainoflayer}, and multidimensional taxonomy adaptation for evolving research corpora~\cite{kargupta2025taxoadapt}. 
Our method improves upon this LLM-based line of work, which expands taxonomies branch by branch and hence produces trees: trees cannot express the many-to-many paper-to-topic relation or multi-parent topics, and branch-local expansion leaves equivalent topics duplicated across branches at incomparable granularities, distorting the downstream session assignment. \sys{} instead canonicalizes every proposed topic against a global registry, collapsing equivalent topics onto shared multi-parent nodes of a conference-specific topic DAG.

%% file: sections/sec_9_conclusion.tex
\section{Conclusion}
We presented \sys{}, a taxonomy-guided system for automatic conference program generation. \sys{} builds a canonicalized multi-parent taxonomy over accepted papers, derives a taxonomy-aware paper distance from it, and solves an integer programming problem that groups papers into coherent sessions under operational constraints. Across four conference benchmarks, \sys{} produced the most coherent sessions and the closest agreement with human-curated programs without constraint violations, and its program for SIGIR~2026 was deployed after only minor organizer edits.

%% file: sections/sec_appendix.tex
\appendix

\section{Details of the Exact and Fallback Solution}
\label{app:solving}

\noindent\textbf{Soft-constraint reformulation.}
Each relaxed constraint is augmented with a nonnegative slack variable $\xi_i$ that measures its degree of violation, and the slacks are penalized in the objective:
\begin{equation}
    \min \sum_{s\in\Sessions}
    \sum_{\substack{p,q\in\Papers}}
    \omega_s d_{pq} z_{pqs}
    \;+\;
    \sum_{i}\lambda_i \xi_i,
    \qquad \xi_i \ge 0.
    \label{eq:soft-relaxation}
\end{equation}
For example, relaxing the lower bound in Constraint~\eqref{eq:capacity} to $\ell_s y_s - \xi_s \le \sum_{p}x_{ps}$ permits an undersized session at a penalty of $\lambda_s\xi_s$, and relaxing a placement restriction permits an out-of-preference assignment at a corresponding cost. The penalty weights $\lambda_i$ encode the organizers' priorities: constraints they regard as more important receive larger weights, and when every $\lambda_i$ dominates the coherence term, the solution minimizes constraint violations lexicographically before optimizing coherence. Assignment completeness~\eqref{eq:assign-once}, presenter conflicts~\eqref{eq:presenter-conflict}, and the binary domains~\eqref{eq:binary-domain} are never relaxed, since violating them would not yield a valid conference schedule; the relaxed problem is therefore always feasible once every paper has at least one admissible session. After solving, the slack values report which soft constraints were violated and by how much, so organizers can either accept the relaxed schedule or revise the session inventory.

\noindent\textbf{Time complexity.}
After the linearization in Constraint~\eqref{eq:z-linear}, the formulation has $O(|\Papers|\,|\Sessions|)$ assignment variables and $O(|\Papers|^2|\Sessions|)$ co-assignment variables, and the underlying problem generalizes capacitated clustering and is NP-hard, so branch-and-bound is exponential in the worst case~\cite{wolsey2020integer}. Two structural properties keep the practical instances small. First, track and placement compatibility fix most assignment variables to zero before the solver is invoked, since a paper can only enter sessions of its own track and admissible slots. Second, the co-assignment terms decompose along tracks: papers in different tracks never share a session, so the problem splits into independent per-track subproblems, each with hundreds of papers and tens of sessions rather than the full conference. As a result, the instances arising in conference organization are solved to optimality, or to a small certified optimality gap under a time limit, within practical time; Sec.~\ref{sec:efficiency} reports the measured running times and costs.

\section{Details of Board Placement for the Poster Program}
\label{app:board-placement}
This appendix details the board-placement formulation and procedure summarized in Section~\ref{sec:post-organization-finalization}. Placement accounts for the two factors introduced there: topical proximity, encoded in an affinity-weighted linear arrangement objective solved by spectral seriation, and presenter proximity, enforced through optional must-link constraints.

\noindent\textbf{Notation and formulation.}
Let $\Papers^{\mathrm{p}}$ be the poster papers, including demonstrations, let $\mathcal{Q}$ be the set of poster sessions, and let $A_p$ be the presenter set of paper $p$. Within each session $q$ with posters $\Papers^{\mathrm{p}}_q=\{p:y_{pq}=1\}$, boards are indexed $1,\dots,B_q$ along the linear traversal of the physical layout, with a powered subset $\mathcal{B}^{\mathrm{pw}}_q$. A partial permutation $z_{pb}\in\{0,1\}$ assigns poster $p$ to board $b$, with $\sum_{b}z_{pb}=1$ and $\sum_{p}z_{pb}\le1$, and $\pi_p=\sum_{b}b\,z_{pb}$ is the board index of $p$. Topic proximity between two posters is measured by their taxonomy affinity $\sigma^{\mathrm{top}}_{pp'}=1/(1+W_1(\mu_p,\mu_{p'};\delta_{\mathcal{G}}))$, and \sys{} minimizes the affinity-weighted linear arrangement
\begin{equation}
    \min\sum_{\substack{p,p'\in\Papers^{\mathrm{p}}_q\\p<p'}}\sigma^{\mathrm{top}}_{pp'}\,\lvert\pi_p-\pi_{p'}\rvert,
    \label{eq:board-mla}
\end{equation}
so that high-affinity posters land on nearby boards.

\noindent\textbf{Spectral seriation.}
Objective~\eqref{eq:board-mla} is a weighted minimum linear arrangement, which generalizes the unweighted minimum linear arrangement (also called optimal linear arrangement), an NP-complete problem~\cite{garey1976simplified}. Exact optimization is therefore expensive in general, so \sys{} uses the objective as a design criterion and computes a scalable spectral-seriation ordering~\cite{atkins1998spectral,fogel2016spectral}. For each session $q$, \sys{} forms the topic-affinity matrix $\Sigma_q=[\sigma^{\mathrm{top}}_{pp'}]$, constructs its graph Laplacian $L_q=D_q-\Sigma_q$, and sorts posters by the Fiedler vector of $L_q$. This ordering places posters with similar affinity profiles nearby in the sequence, which is consistent with the goal of concentrating high-affinity pairs along nearby board positions.

\noindent\textbf{Presenter proximity and powered boards.}
When presenter proximity is enabled, \sys{} first constructs disjoint must-link groups, for example the connected components induced by shared presenters and demonstration--companion relations. Each group is contracted into a single unit before spectral seriation, with inter-unit affinities obtained by aggregating the pairwise affinities between their member posters. After the units are ordered, each contracted group is expanded into a contiguous block, and the members inside the block are ordered by the same topic-affinity criterion. This procedure enforces presenter-level co-location while preserving the topic-driven ordering among the remaining posters. Powered-board constraints for demonstrations are then enforced during block placement. If no powered board in $\mathcal{B}^{\mathrm{pw}}_q$ is compatible with a required demonstration block, \sys{} reports the conflict to the organizer or relaxes the optional presenter-proximity constraint. Taxonomy thus determines which posters should be near each other, while the arrangement step converts this preference into a concrete, walkable layout.

\section{Implementation Details}
\label{sec:impl-details}

\subsection{Benchmark Construction}
Each benchmark is derived from the official program of its venue. The human-curated sessions define the candidate inventory $\Sessions$ given to every method: the number of sessions, their time slots, and a venue-wide capacity range set to the minimum and maximum observed session sizes (Table~\ref{tab:dataset_statistics}). The human paper-to-session grouping is held out as the reference partition, and session names are never shown to any method. Presenter-conflict pairs are derived from normalized author lists, and paper identifiers are deterministically shuffled so that identifier adjacency cannot leak the human grouping to the LLM-based methods. All methods receive titles and abstracts only.

\subsection{\sys{} Configuration}
Taxonomy construction (Alg.~\ref{alg:taxonomy-construction}) uses minimum node size $\theta_{\min}{=}5$, maximum depth $L_{\max}{=}4$, and at most 10 proposed children per expansion; canonicalization retrieves the top-8 registry candidates by embedding similarity and resolves equivalence with a single LLM call, and routing returns one Boolean per child under a strict JSON schema. The topic distance $\delta_{\mathcal{G}}$ is computed with Dijkstra's algorithm, and the paper distance $d^{\mathrm{tax}}_{pq}$ is solved exactly as a transportation LP with memoization over distinct frontiers. The assignment problem is solved with CP-SAT (OR-Tools) using 8 workers, a fixed seed, and a 300\,s limit per instance, warm-started from a feasible incumbent built by agglomerative grouping and conflict-aware repair; the soft-constraint fallback of Eq.~\eqref{eq:soft-relaxation} never triggers on these benchmarks. All LLM components use \texttt{gpt-5.4-mini}~\cite{openai2026gpt54mini} with strict JSON outputs.

\subsection{Baselines}
All baselines share the benchmark loader and session inventory, and the ILP variants additionally share the assignment formulation and solver budget above, so only the distance $d_{pq}$ changes: \TopicILP{} uses the Jensen--Shannon distance between LDA topic distributions with $K$ equal to the number of sessions, and \EmbILP{} uses cosine distance between SPECTER2 embeddings. \LLMDirect{} emits the full program in a single generation under a strict JSON schema, with no optimization or repair; only well-formedness is patched, and every patch is logged so that capacity and conflict violations remain visible to the evaluation. \LLMReAct{} runs at most three propose--check--revise rounds, combining a programmatic constraint report with an LLM coherence critique. Both LLM-based baselines use the same \texttt{gpt-5.4-mini} backbone as \sys{} to ensure a fair comparison.

\subsection{Evaluation}
For the LLM-based coherence evaluation, we employ a panel of state-of-the-art LLM judges from three model families, \texttt{gpt-5.5}~\cite{openai2026gpt55}, \texttt{claude-opus-4.8}~\cite{anthropic2026claudeopus48}, and \texttt{gemini-3.1-pro}~\cite{google2026gemini31pro},  and report the average of their scores. Every judge call is cached on disk keyed by its full request content, so the evaluation is deterministic and re-runnable.

\section{Ablation Study}
\label{app:ablation}

\input{tables/ablation}

We ablate each major component of \sys{} while holding the rest of the pipeline fixed. Every variant solves the same constraint-aware integer programming formulation under the same solver budget as the full system, so all differences are attributable to the semantic representation that produces the pairwise distances $d_{pq}$. All variants are evaluated on the four benchmarks under the protocol of Sec.~\ref{sec:main_results}, and Table~\ref{tab:ablation} reports the results averaged over the benchmarks.

\begin{itemize}[leftmargin=*,noitemsep,topsep=0pt]
\item \textbf{w/o canonicalization.} The global registry is disabled: every topic proposed during expansion becomes a new node under its proposing parent. Equivalent topics discovered in different branches remain distinct, and the taxonomy degenerates to a branch-local tree. This variant measures the contribution of global topic deduplication.
\item \textbf{w/o multi-parent routing.} Canonicalization is retained, but each canonical topic keeps only its first discovered parent, so the taxonomy is a deduplicated tree. Cross-cutting topics lose their secondary specialization edges. This variant isolates the contribution of the DAG structure from that of deduplication.
\item \textbf{w/o specificity weighting.} The specificity-gain edge lengths are replaced by unit lengths, so the topic distance reduces to an undirected hop count and reflects structural depth rather than topic specificity.
\item \textbf{w/o optimal transport.} The optimal-transport paper distance is replaced by single-linkage matching, taking the minimum topic distance between the two frontiers. This variant measures the benefit of comparing full frontier distributions rather than closest topic pairs.
\item \textbf{w/o frontier representation.} Each paper is represented by all of its assigned topics $A(p)$ with uniform mass instead of its frontier $F(p)$, so broad ancestors dilute the most specific topics.
\end{itemize}

Removing any single component degrades every quality metric, while feasibility is unaffected: all variants retain zero violations because the hard constraints are enforced by the integer program independently of the distances. The components that shape the paper distance contribute most. Replacing the frontier representation with all assigned topics causes the largest degradation (NMI drops from 0.761 to 0.692 and the outlier count doubles from 12.8 to 25.6), confirming that diluting a paper's most specific topics with broad ancestors blurs exactly the distinctions that sessions are built from. Replacing optimal transport with single-linkage matching is the second most damaging (NMI 0.705), since the minimum topic distance ignores all but the closest facet of multi-topic papers, and removing specificity weighting follows (NMI 0.718), as hop counts conflate structural depth with topic specificity across unevenly refined branches. The taxonomy-structure ablations are milder but consistent: retaining a deduplicated tree without multi-parent routing costs 0.016 NMI, and disabling canonicalization altogether costs 0.029, indicating that global deduplication and cross-branch specialization edges each contribute. Notably, even the weakest variant remains above the strongest optimization baseline of Table~\ref{tab:main_results} on every agreement metric (NMI 0.692 vs.\ 0.681 for \EmbILP{} on average) and clearly above it on coherence (4.41 vs.\ 4.14), so the taxonomy-based distance is beneficial even in reduced form, and the full margin of \sys{} comes from the accumulation of all five components.

\section{Attendee Survey at SIGIR~2026}
\label{sec:attendee-survey}

\begin{figure}[t]
    \centering
    \includegraphics[width=\linewidth]{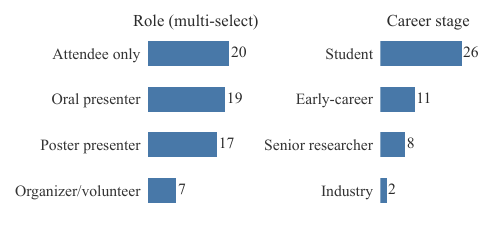}
    \caption{Respondent distribution of the SIGIR~2026 attendee survey ($N{=}47$). Role is multi-select, so role counts sum to more than $N$.}
    \label{fig:survey_respondents}
\end{figure}

\begin{figure}[t]
    \centering
    \includegraphics[width=\linewidth]{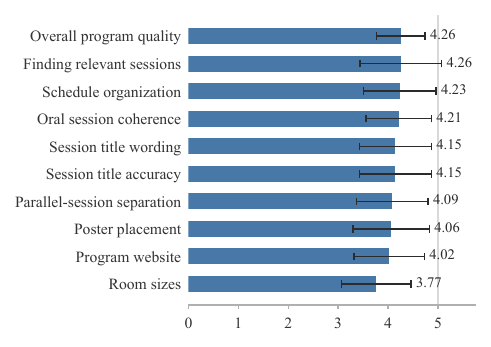}
    \caption{Average scores for the ten surveyed aspects of the SIGIR~2026 program (5-point scale); error bars show one standard deviation.}
    \label{fig:survey_scores}
\end{figure}

At SIGIR~2026, we conducted an anonymous on-site survey\footnote{\url{https://forms.gle/mjY4hKfyZtq5ZauA8}} of participants to assess the perceived quality of the deployed program. Attendees were recruited in two ways at the conference venue: (1)~QR codes linking to the survey were posted on site, and (2)~we approached attendees of different backgrounds in person, introduced how the program had been generated, and invited them to complete the survey. We received 47 responses. Respondents reported their roles (multi-select) and career stages, rated ten aspects of the program on a 5-point scale, and could leave optional free-text comments on what they liked, what should be improved, and what the online program website should add. Figure~\ref{fig:survey_respondents} shows the respondent distribution: the sample covers attendees (20), oral presenters (19), poster presenters (17), and organizers or volunteers (7), and spans students (26), early-career researchers (11), senior researchers (8), and industry practitioners (2).

Figure~\ref{fig:survey_scores} reports the average score of each aspect together with its standard deviation. Every aspect has a median of 4, with means ranging from 3.77 to 4.26 and standard deviations between 0.49 and 0.82, the tightest agreement being on overall program quality (s.d.\ 0.49). The aspects that reflect \sys{}'s assignment quality all score well: overall program quality and ease of finding relevant sessions score highest (4.26), followed by schedule organization (4.23), oral session coherence (4.21), session title accuracy and wording (both 4.15), parallel-session separation (4.09), and poster placement (4.06). The lowest-rated aspect is the adequacy of room sizes (3.77); consistently, 12 of the 47 respondents mentioned crowded rooms for popular sessions in the free-text comments, making popularity-based room allocation the clearest target for improvement. The remaining free-text suggestions concern the online program website (most commonly saved-session bookmarks, more visible mobile filters, and calendar export) and occasional requests for more specific session titles; these are orthogonal to session assignment and inform future iterations of the system.

%% file: tables/ablation.tex
\begin{table}[t]
\centering
\footnotesize
\setlength{\tabcolsep}{2.5pt}
\renewcommand{\arraystretch}{1.0}
\caption{Ablation results averaged over the four benchmarks. Each variant disables one component of TaxoConf; all variants solve the same constraint-aware integer programming formulation under the same solver budget. Arrows give the preferred direction. The best result is in \textbf{bold}.}
\label{tab:ablation}
\begin{tabular}{@{}l cc ccc cc@{}}
\toprule
\multirow{2}{*}{\textbf{Variant}} & \multicolumn{2}{c}{\textbf{LLM Eval.}} & \multicolumn{3}{c}{\textbf{Human Agreement}} & \multicolumn{2}{c}{\textbf{Violation}} \\
\cmidrule(lr){2-3}\cmidrule(lr){4-6}\cmidrule(lr){7-8}
& \textbf{Coh.}$\uparrow$ & \textbf{Out.}$\downarrow$ & \textbf{NMI}$\uparrow$ & \textbf{ARI}$\uparrow$ & \textbf{F1}$\uparrow$ & \textbf{Cap.}$\downarrow$ & \textbf{Conf.}$\downarrow$ \\
\midrule
\textbf{TaxoConf (full)}       & \textbf{4.62} & \textbf{12.8} & \textbf{0.761} & \textbf{0.242} & \textbf{0.541} & \textbf{0} & \textbf{0} \\
\midrule
w/o canonicalization         & 4.54 & 16.5 & 0.732 & 0.215 & 0.512 & 0 & 0 \\
w/o multi-parent routing     & 4.58 & 14.2 & 0.745 & 0.231 & 0.528 & 0 & 0 \\
w/o specificity weighting    & 4.49 & 18.7 & 0.718 & 0.198 & 0.495 & 0 & 0 \\
w/o optimal transport        & 4.45 & 21.3 & 0.705 & 0.185 & 0.472 & 0 & 0 \\
w/o frontier representation  & 4.41 & 25.6 & 0.692 & 0.170 & 0.455 & 0 & 0 \\
\bottomrule
\end{tabular}
\end{table}